\documentclass[prb,twocolumn,showpacs,preprintnumbers,amsmath,amssymb]{revtex4}
\usepackage{graphicx}
\newcommand{\tvec}[2]{\left( \begin{array}{c}
   \!\! \mbox{$#1$} \!\! \\ \!\! \mbox{$#2$} \!\! \end{array} \right)}

\begin{document}

\title{Andreev Reflection in Ferromagnet/Superconductor/Ferromagnet
  Double Junction Systems}
\author{Taro Yamashita$^1$, Hiroshi Imamura$^2$,
Saburo Takahashi$^1$, and Sadamichi Maekawa$^1$}
\affiliation{
$^{1}$Institute for Materials Research, Tohoku University, Sendai 980-8577, Japan\\
$^{2}$Graduate School of Information Sciences, Tohoku University, Sendai 980-8579, Japan}

\date{\today}

\begin{abstract}
  We present a theory of Andreev reflection in
  a ferromagnet/superconductor/ferromagnet double junction system.  
  The spin polarized quasiparticles penetrate to the superconductor in
  the range of penetration depth from the interface
  by the Andreev reflection. 
  When the thickness of the superconductor is comparable
  to or smaller than the penetration depth, the spin polarized
  quasiparticles pass through the superconductor and therefore the electric
  current depends on the relative orientation of magnetizations of the
  ferromagnets.  The dependences of the magnetoresistance on
  the thickness of the superconductor, temperature, the exchange field
  of the ferromagnets and the height of the interfacial barriers are analyzed.
  Our theory explains recent experimental results well.
\end{abstract}

\maketitle

\section{Introduction}\label{sec:intro}
The spin-dependent transport through magnetic
nanostructures has attracted much interest.\cite{maekawaBOOK}  
In the early 1970s, Meservey and Tedrow have showed that tunneling electrons
between a ferromagnetic metal (Fe, Co, Ni)
and a thin film of superconducting aluminium
(Al) are spin-polarized.\cite{meservey}  
The ferromagnet/insulator/superconductor (FM/I/SC) tunnel junctions are one of
the most powerful tools to extract the
spin polarization of the conduction electrons near the Fermi level.  
In FM/I/SC and FM/I/SC/I/FM tunnel junctions,
the suppression of superconducting gap due to spin accumulation
by injection of spin polarized quasiparticles
(QP's) has been shown.\cite{vasko,dong,yeh,liu, chen,takahashiPRL}  
The QP's spin transport and relaxation in SC has been studied in detail.
\cite{yamashita,takahashiJMMM,takahashiHOLL}

In recent years, much attention has been focused on FM/SC metallic
contacts both
theoretically\cite{de_jong,kikuchi,hima,zhu,igor,kashiwaya}
and experimentally \cite{soulen,soulen2,ji,strijkers,shashi} since
the spin polarization of conduction electrons is measured by using the
Andreev reflection\cite{andreev}:  An electron injected from the FM into the SC
is reflected as a hole
at the FM/SC interface and a Cooper pair is generated in the SC.  
The Andreev reflection includes a conversion process of the QP current
to the supercurrent carried by Cooper pairs in the range of the
penetration depth, which is approximately equal to the Ginzburg-Landau (GL)
coherence length,\cite{GL} from the FM/SC interface.\cite{BTK}  
Thus, a FM/SC/FM double junction is particularly interesting because
the magnetoresistance is expected due to the overlap of the QP penetration
in the SC by the Andreev reflection.  
Recently, Gu {\it et al.} have measured the magnetoresistance
in a current perpendicular to plane (CPP) geometry consisting of
a superconducting niobium (Nb) thin film
sandwiched by ferromagnetic permalloys (Py) and
proposed a method for estimating the penetration depth
by measuring the magnetoresistance.\cite{gu}  



In this paper, we present a theory of the Andreev
reflection in the FM/SC/FM double junction system and derive an expression of
the current through the junction by extending the theory of Blonder, Tinkham and Klapwijk (BTK).\cite{BTK}  
We numerically calculate the current for the parallel and anti-parallel alignments of magnetizations, and investigate the dependences of the
magnetoresistance on the thickness of the SC, temperature, the
exchange field of FM's and the height of the interfacial
barriers.  It is shown that these dependences are understood by considering
the penetration of quasiparticles to the SC by the Andreev
reflection process.  
Finally, we compare our results with the recent experimental results by
Gu {\it et al.}.\cite{gu}



\section{Model and Formulation}\label{sec:model}

\begin{figure}
\includegraphics[width=\columnwidth]{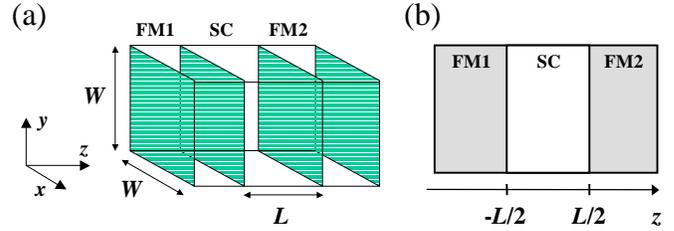}
\caption{(a) Schematic diagram of a
  ferromagnet/superconductor/ferromagnet (FM1/SC/FM2) double junction
  system.  A superconductor with a thickness of $L$ is sandwiched by two
  semi-infinite ferromagnetic electrodes.  The system is rectangular
  and the cross section is a square of side $W$.
  (b) The current flows along the $z$-axis.  The interfaces between
  FM1/SC and SC/FM2 are located at $z=-L/2$ and $z=L/2$,
  respectively. }
\label{fig:geo}
\end{figure}
We consider a FM1/SC/FM2 double junction system consisting of three
rectangular blocks as shown in Figs. 1(a) and 1(b).  The cross section
of the system is a square of side $W$ and 
the thickness of the SC is  $L$.  The current
flows along the $z$-axis and the interfaces between FM1/SC and
SC/FM2 are located at $z=-L/2$ and $z=L/2$, respectively.
For simplicity, we assume that the system is symmetric: FM1 and FM2
are made of the same ferromagnetic materials and the potentials for
the left and right interfaces are the same.
The system we consider is described by the following Bogoliubov-de
Gennes (BdG) equation \cite{BdG}:
\begin{eqnarray}
\left( {\begin{array}{*{20}c}\!\!
   {H_0 \! - \! h_{ex}(z)\sigma} & \Delta (z)  \\
   {\Delta ^* (z)} & { - {H_0 \! - \! h_{ex}(z)\sigma}}  \\
\end{array}} \!\! \right) \!\! \left( \!\! {\begin{array}{*{20}c}
   {f_{\sigma}\left( {{\bf r}} \right)}  \\
   {g_{\sigma}\left( {{\bf r}} \right)}  \\
\end{array}} \!\! \right) \!
= \! E \! \left( \!\! {\begin{array}{*{20}c}
   {f_{\sigma}\left( {{\bf r}} \right)}  \\
   {g_{\sigma}\left( {{\bf r}} \right)}  \\
\end{array}} \!\! \right) ,
\label{eq:BdG}
\end{eqnarray}
where $H_0 \equiv -({\hbar}^2/2m){\nabla}^2-\mu_F$ is the single
particle Hamiltonian, $E$ is the QP energy measured from the Fermi
energy $\mu_F$ and $\sigma=+(-)$ is for the up-(down-)spin band.  
The exchange field $h_{ex}\left( z \right)$ is given by 
\begin{equation}
h_{ex}\left( z \right) = \left\{ 
\begin{array}{ll}
h_{0} & (z < -L/2) \\
0      & (-L/2 < z < L/2) \\
\pm h_{0} & (L/2 < z)
\end{array} \right. ,
\end{equation}
where $+h_{0}$ and $-h_{0}$ represent the exchange fields for the parallel
and anti-parallel alignments, respectively.
The superconducting gap is expressed as
\begin{equation}
\Delta \left( z \right) 
= \left\{ 
  {\begin{array}{cl}
      0      & (z < -L/2, L/2 < z) \\
      \Delta & (-L/2 < z < L/2)
    \end{array}} \right. .
\end{equation}
We assume that the temperature dependence of the superconducting
gap is given by $\Delta = \Delta_0
\tanh{\left( 1.74\sqrt{{T_c}/{T}-1}\right)}$,\cite{belzig} where $\Delta_0$ is
the superconducting gap at $T=0$ and $T_c$ is the superconducting
critical temperature.  
In order to capture the essential effect of the interfacial
scattering, we employ the following $\delta$-function type potential
at the interfaces:
\begin{equation}
H\left( z \right) = \frac{\hbar^2k_F}{m}Z
 \left\{\delta \left( z + L/2 \right) 
+  \delta \left( {z - L/2} \right) \right\}.  
\end{equation}
Throughout this paper, we neglect the spin-flip scattering in the SC
and the proximity effect near the interfaces.\cite{strijkers}  

Since the system is rectangular, the wave function in the
transverse ($x$ and $y$) directions is given by 
\begin{equation}
{\rm S}_{nl}(x,y) \equiv \sin{\left( n\pi x/W
  \right)}\sin{\left( l\pi y/W \right)},
\label{eq:snl}
\end{equation}
where $n$ and $l$ are the quantum numbers which define the channel.
The eigenvalue of the transverse mode for the channel ($n$,$l$) is
\begin{eqnarray}
  E_{nl} = \frac{\hbar^2}{2m}\left[ {{\left( \frac{n\pi}{W}\right)}^2 + {\left( \frac{l\pi}{W}\right)}^2 }\right].  
\end{eqnarray}

The solution of the BdG equation (\ref{eq:BdG}) in the
SC region is given by
\begin{eqnarray}
\begin{array}{l}
 \Psi _{ \pm k_{nl}^+ } ({\bf r}) 
 = \tvec{u_0}{v_0 }
 e^{ \pm ik_{nl}^+ z} \, {\rm S}_{nl}(x,y),\\
 \Psi _{ \pm k_{nl}^- } ({\bf r}) 
 = \tvec{v_0 }{u_0 }
 e^{ \pm ik_{nl}^- z} \, {\rm S}_{nl}(x,y) ,
\end{array}
\label{eq:wf-SC}
\end{eqnarray}
where $u_0$ and $v_0$ are the coherence factors expressed as
\begin{eqnarray}
u_0^2  = 1 - v_0^2  = \frac{1}{2}\left[ {1 + \frac{{\sqrt {E^2  - \Delta ^2 } }}{E}} \right] ,
\end{eqnarray}
and $k_{nl}^{+(-)}$ is the $z$ component of the wave number
of an electron-(hole-)like QP in the channel $(n,l)$ defined as
\begin{eqnarray}
k_{nl}^{\pm} = \frac{\sqrt{2m}}{\hbar}  \sqrt{\mu _F \pm \sqrt{E^2 - \Delta^2
    }-E_{nl}}.
\end{eqnarray}

In the FM region, the solutions are given by
\begin{eqnarray}
\begin{array}{l}
 \Psi _{ \pm p_{\sigma ,nl}^+ } ({\bf r}) 
 = \tvec{1}{0} 
 e^{ \pm ip_{\sigma ,nl}^+ z} \, {\rm S}_{nl}(x,y), \\
 \Psi _{ \pm p_{\sigma ,nl}^- } ({\bf r}) 
 = \tvec{0}{1}
 e^{ \pm ip_{\sigma ,nl}^- z} \, {\rm S}_{nl}(x,y), 
\end{array}
\label{eq:wf-FM}
\end{eqnarray}
where $p_{\sigma ,nl}^{+(-)}$ is the $z$ component of the wave number
of an electron (hole) with $\sigma$-spin in the channel $(n,l)$;
\begin{eqnarray}
 p_{\sigma ,nl}^{\pm} 
= \frac{\sqrt{2m}}{\hbar} \sqrt{\mu _F \pm E \pm \sigma h_{ex} - E_{nl}}.
\end{eqnarray}
\begin{figure}
\includegraphics[width=\columnwidth]{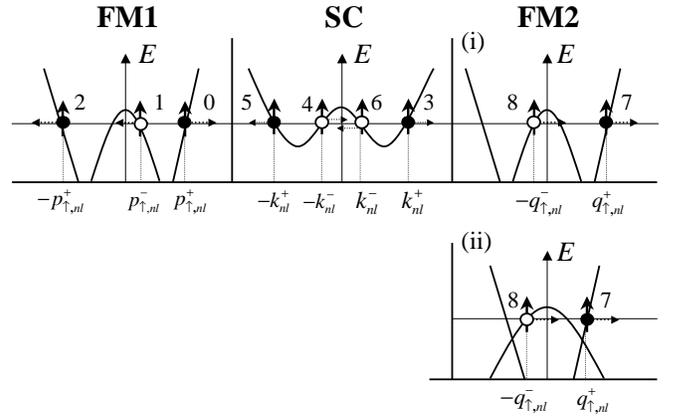}
\caption{Schematic diagrams of energy vs. momentum for the FM1/SC/FM2 double
  junction system with the parallel and anti-parallel alignments of the
  magnetizations are shown in panels (i) and (ii), respectively.  
  The open circles denote holes, the closed circles
  electrons, and the arrows point in the direction of the group
  velocity.  
  The incident electron with up-spin in the channel ($n,l$) is denoted by 0,
  along with the resulting scattering processes: the Andreev
  reflection (1), the normal reflection (2) at the interface of
  FM1/SC, the transmission to the SC (3, 4) and the reflection at the
  interface of SC/FM2 (5, 6), the transmission as an electron to the FM2
  (7) and the one as a hole (8).}
\label{fig:process}
\end{figure}
The wave function of the FM1/SC/FM2 double junction system is given by
the linear combination of the solutions.
Let us consider the scattering of an electron with up-spin in the
channel $(n,l)$ injected into the SC from the FM1 (0 in Fig.\ref{fig:process}).
There are the following eight processes: the Andreev reflection (1 in
Fig.\ref{fig:process}), the normal reflection (2 in
Fig.\ref{fig:process}) at the interface of FM1/SC, the transmission to
the SC (3, 4 in Fig.\ref{fig:process}) and the reflection at the
interface of SC/FM2 (5, 6 in Fig.\ref{fig:process}), the transmission
as an electron to the FM2 (7 in Fig.\ref{fig:process}) and the one as
a hole (8 in Fig.\ref{fig:process}).  
Therefore, the wave function in the FM1 $(z<-L/2)$ is given by 
\begin{eqnarray}
\Psi ^{\rm FM1}_{\sigma ,nl}({\bf r})
 &=& \Bigg[ 
 \tvec{1}{0}
 e^{ip_{\sigma ,nl}^{+}\left( z+\frac{L}{2}\right) }
 + a_{\sigma ,nl} 
 \tvec{0}{1}
 e^{ip_{\sigma ,nl}^{-}\left( z+\frac{L}{2}\right) }
\nonumber \\
 &+& b_{\sigma, nl}
 \tvec{1}{0}
 e^{-ip_{\sigma ,nl}^{+}\left( z+\frac{L}{2}\right) }
\Bigg] {\rm S}_{nl}(x,y).
\label{eq:wf2-FM1}
\end{eqnarray}
In the SC $(-L/2<z<L/2)$ we have
\begin{eqnarray}
&&\Psi ^{\rm SC}_{\sigma ,nl}({\bf r}) = \Bigg[
\alpha_{\sigma ,nl} 
\tvec{u_0}{v_0}
e^{ik_{nl}^+ \left( z+\frac{L}{2}\right) } \nonumber\\
&&+ \beta_{\sigma ,nl} 
\tvec{v_0}{u_0}
e^{-ik_{nl}^- \left( z+\frac{L}{2}\right) }
+ \xi_{\sigma ,nl} 
\tvec{u_0}{v_0}
e^{-ik_{nl}^+ \left( z-\frac{L}{2}\right) } \nonumber\\
&&+ \eta_{\sigma ,nl} 
\tvec{v_0}{u_0 }
e^{ik_{nl}^- \left( z-\frac{L}{2}\right) }\Bigg]
   {\rm S}_{nl}(x,y),
\label{eq:wf2-SC}
\end{eqnarray}
and in the FM2 $(L/2<z)$
\begin{equation}
\begin{split}
\Psi ^{\rm FM2}_{\sigma ,nl}({\bf r}) 
&= \Bigg[
  c_{\sigma ,nl} 
  \tvec{1}{0}
  e^{iq_{\sigma ,nl}^{+}\left( z-\frac{L}{2}\right) }\\
&+ d_{\sigma, nl}
\tvec{0}{1}
e^{-iq_{\sigma ,nl}^{-}\left( z-\frac{L}{2}\right) }
\Bigg] {\rm S}_{nl}(x,y).
\end{split}
\label{eq:wf2-FM2}
\end{equation}
Here $p_{\sigma ,nl}^{\pm}$, $k_{nl}^{\pm}$ and $q_{\sigma
  ,nl}^{\pm}$ are the wave numbers in the FM1, SC and FM2, respectively.
The coefficients $a_{\sigma, nl},
b_{\sigma, nl}, c_{\sigma, nl}, d_{\sigma, nl}, \alpha_{\sigma, nl},
\beta_{\sigma, nl}$, $\xi_{\sigma, nl}$, and $\eta_{\sigma, nl}$ are
determined by matching the wave functions at the left and right interfaces. 
The matching conditions for the wave functions (\ref{eq:wf2-FM1}) -
(\ref{eq:wf2-FM2}) at the interfaces are 
\begin{equation}
  \left\{
    \begin{split}
& \Psi ^{\rm FM1}_{\sigma ,nl} \left( \mbox{$z=-\frac{L}{2}$} \right)
 =\Psi ^{\rm SC}_{\sigma ,nl}\left( \mbox{$z=-\frac{L}{2}$} \right),\\
& \Psi ^{\rm SC}_{\sigma ,nl} \left( \mbox{$z=\frac{L}{2}$} \right) = \Psi ^{\rm FM2}_{\sigma ,nl} \left( \mbox{$z=\frac{L}{2}$} \right),\\
&\frac{d\Psi ^{\rm SC}_{\sigma ,nl}}{{dz}}\Bigr|_{z=-\frac{L}{2}} 
\!\!- \frac{d\Psi ^{\rm FM1}_{\sigma
    ,nl}}{{dz}}\Bigr|_{z=-\frac{L}{2}}
\!=\! \frac{{2m Z }}{{\hbar^2 }}\Psi ^{\rm FM1}_{\sigma ,nl} \left( \mbox{$z=-\frac{L}{2}$} \right),\\
&\frac{d\Psi ^{\rm FM2}_{\sigma ,nl}}{{dz}}\Big|_{z=\frac{L}{2}} 
\!\!- \frac{d\Psi ^{\rm SC}_{\sigma ,nl}}{{dz}}\Big|_{z=\frac{L}{2}}
\!=\! \frac{{2m Z }}{{\hbar^2 }}\Psi ^{\rm FM2}_{\sigma ,nl} \left(\mbox{$z=\frac{L}{2}$} \right) .
\end{split}
\right.
\label{eq:bound}
\end{equation}
Solving Eq.(\ref{eq:bound}), the probabilities of
transmission and reflection are calculated following the BTK theory.
\cite{BTK}  When an electron with $\sigma$-spin is injected from the FM1,
the probability of the Andreev reflection $R_{\sigma
  ,nl}^{he}$, the normal reflection $R_{\sigma ,nl}^{ee}$ and the
transmission as an electron and the one as a hole to the FM2, $T_{\sigma
  ,nl}^{e'e}$ and $T_{\sigma ,nl}^{h'e}$ are given by
\begin{equation}
\left\{
\begin{split}
&R_{\sigma ,nl}^{he} \left( E \right) = {\frac{p_{\sigma ,nl}^- }{p_{\sigma ,nl}^+ }} \, a_{\sigma ,nl}^* a_{\sigma ,nl},\\
&R_{\sigma ,nl}^{ee} \left( E \right) = b_{\sigma ,nl}^* b_{\sigma ,nl},\\
&T_{\sigma ,nl}^{e'e} \left( E \right) = {\frac{q_{\sigma ,nl}^+ }{p_{\sigma ,nl}^+ }} \, c_{\sigma ,nl}^* c_{\sigma ,nl},\\
&T_{\sigma ,nl}^{h'e} \left( E \right) = {\frac{q_{\sigma ,nl}^- }{p_{\sigma ,nl}^+ }} \, d_{\sigma ,nl}^* d_{\sigma ,nl} ,
\end{split}
\right.
\label{eq:rt}
\end{equation}
where the subscript $e'(h')$ in Eq. (\ref{eq:rt})
indicates the electron (hole) in the FM2.
Let us evaluate the current in the FM1.
When the bias voltage $V$ is applied to the FM1/SC/FM2 system, the current
carried by electrons with $\sigma$-spin is written as
\begin{eqnarray}
I_{\sigma}^{e}   = \frac{e}{h}\sum_{nl}\int_0^\infty  {\left[ {f_\to \left( E \right) - f_\leftarrow \left( E \right)} \right]} dE ,
\end{eqnarray}
where $h$ is Planck constant and $f_{\to} \left( E \right)$ is the
distribution function of an electron with a positive group velocity in
the $z$ direction and expressed as
\begin{equation}
f_\to \left( E \right) = f_0 \left( \mbox{$E-\frac{{eV}}{2}$} \right) ,
\end{equation}
where $f_0 \left( E \right)$ is the Fermi distribution function.  The
distribution function of the electron with a negative group velocity
in the $z$ direction is written as
\begin{eqnarray}
&&f_\leftarrow \left( E \right)\nonumber\\
&&= R_{\sigma,nl}^{eh} f_0 \left( \mbox{$E + \frac{eV}{2}$} \right) +
R_{\sigma,nl}^{ee} f_0 \left( \mbox{$E - \frac{eV}{2}$} \right)\nonumber\\ 
&&\ \ \ +
\frac{{v_{ \sigma ,nl}^R{\cal D}_{\sigma ,nl}^{R}}}{{v_{ \sigma ,nl}^L
    {\cal D}_{\sigma ,nl}^{L} }} \left[ T_{\sigma,nl}^{ee'} f_0 \left
    ( \mbox{$E + \frac{{eV}}{2}$} \right) + T_{\sigma,nl}^{eh'} f_0 \left
    ( \mbox{$E - \frac{eV}{2}$} \right) \right]\nonumber\\
&&= R_{\sigma,nl}^{eh} f_0 {\left( \mbox{$E + \frac{eV}{2}$}\right)} +
R_{\sigma,nl}^{ee} f_0 \left( \mbox{$E - \frac{eV}{2}$} \right) \nonumber\\
&&\ \ \ + T_{\sigma,nl}^{ee'} f_0 \left( \mbox{$E + \frac{{eV}}{2}$} \right) +
T_{\sigma,nl}^{eh'} f_0 \left( \mbox{$E - \frac{eV}{2}$} \right),
\end{eqnarray}
where $v_{\sigma ,nl}^{L(R)}$ is the Fermi velocity of an electron with
$\sigma$-spin in the channel $(n,l)$ of the FM1 (FM2) and ${\cal
  D}_{\sigma ,nl}^{L(R)}$ is the density of states of $\sigma$-spin
band in the channel $(n,l)$ of the FM1 (FM2). 
Using the conservation of probability,
$R_{\sigma,nl}^{eh} + R_{\sigma,nl}^{ee} + T_{\sigma,nl}^{ee'} +
T_{\sigma,nl}^{eh'} = 1$, we have
\begin{equation}
\begin{split}
I_{\sigma}^{e} &= \frac{e}{h}\sum_{nl} \int_0^\infty \left(
  { R_{nl,\sigma}^{eh}  + T_{nl,\sigma}^{ee'}} \right)\\
&\times \left[ {f_0 \left( \mbox{$E - \frac{eV}{2}$} \right) - f_0 \left( \mbox{$E + \frac{eV}{2}$} \right) } \right] dE .
\end{split}
\end{equation}
The current carried by holes $I_{\sigma}^{h}$ is calculated in the
similar way.  

The total current in the FM1/SC/FM2 double junction system is
obtained as
\begin{equation}
  \begin{split}
    I &= \sum_{\sigma} \left[ I_{\sigma}^{e} + I_{\sigma}^{h} \right]\\
    &= \frac{e}{h}\sum_{nl,\sigma} \int_0^\infty \left( R_{nl,\sigma}^{eh}
      + R_{nl,\sigma}^{he} + T_{nl,\sigma}^{ee'} + T_{nl,\sigma}^{hh'}
    \right)\\ 
    &\times \left[ {f_0 \left( \mbox{$E - \frac{eV}{2}$} \right) - f_0
        \left( \mbox{$E + \frac{eV}{2}$} \right) }\right]dE .
  \end{split}
\label{current}
\end{equation}
Note that this expression of the current Eq. (\ref{current}) reduces to
the one derived by Lambert\cite{lambert} for
the normal metal/superconductor/normal metal system when $h_{0} = 0$.

The magnetoresistance (MR) is defined as
\begin{equation}
  {\rm MR} \equiv \frac{R_{\rm AP} - R_{\rm P}}{R_{\rm P}},
\end{equation}
where $R_{\rm P(AP)} = V/I_{\rm P(AP)}$ is the resistance in the parallel (anti-parallel) alignment.


\section{Results}

\begin{figure}
\includegraphics[width=\columnwidth]{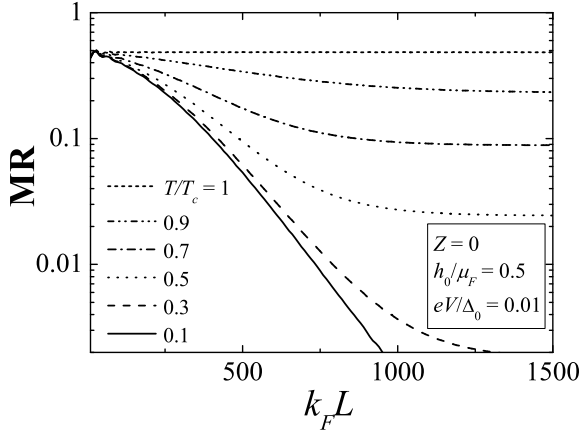}
\caption{MR as a function of the thickness of the SC, $k_{F} L$.  
  From top to bottom, temperature $T/T_c$ is
  1, 0.9, 0.7, 0.5, 0.3, and 0.1.  
  We assume $\xi_{Q}(E=T=0) = 200/k_{F}$.}
\label{fig:Thick}
\end{figure}
In Fig. \ref{fig:Thick} the MR is plotted as a function of
the thickness of the SC, $L$, multiplied by the Fermi wave number $k_{F}$.
We assume that the strength of the interfacial barrier $Z=0$
and the exchange field $h_{0}=0.5\mu_F$.  
The side length of the cross section is taken to be $W=1000/k_{F}$
throughout this paper.  
When the SC is in the normal conducting state ($T/T_c = 1$), 
the MR is constant since we neglect the spin-flip
scattering in the SC.  
When the SC is in the superconducting state ($T/T_c$
= 0.1, 0.3, 0.5, 0.7, and 0.9), the MR decreases with increasing the
thickness of the SC.  The MR at low temperatures $T/T_c \ll 1$
shows an exponential decrease in a wide range of $k_F L$.
The decrease of the MR due to the superconductivity can be explained
by considering the decay of the quasiparticle current in the SC.
To obtain the charge and spin currents, we extend the BTK theory of
the Andreev reflection \cite{BTK} to the FM1/SC/FM2 double junction
system.  

Let us first consider the charge transport in the SC.  
When the current flows in the positive $z$ direction, electrons and holes are injected into the SC from the FM1 and the FM2, respectively.  
The conservation law of the charge density
$Q_{\sigma}=e\sum_{nl}\left( |f_{nl}^{\sigma}|^2
-|g_{nl}^{\sigma}|^2 \right)$ in the SC, where $f_{nl}^{\sigma}$ and
$g_{nl}^{\sigma}$ are electron- and hole-like components of the wave function
in the channel ($n,l,\sigma$), respectively, is derived from the BdG equation
(\ref{eq:BdG}) and obtained as
\begin{equation}
\frac{\partial Q_{\sigma}}{\partial t} + \nabla\cdot {\bf J}_Q^{\sigma} = \frac{4e\Delta}{\hbar}\sum_{nl}{\rm Im}({f_{nl}^{\sigma}}^{\ast} g_{nl}^{\sigma}) ,
\label{contin}
\end{equation}
where ${\bf J}_Q^{\sigma}$ is the QP current density with $\sigma$-spin and the $z$ component of ${\bf J}_Q^{\sigma}$ per unit area $j_{Q}^{\sigma}$ is written as
\begin{eqnarray}
j_{Q}^{\sigma} (z) &=& \frac{e\hbar}{mW^2} \sum_{nl}
\int_{0}^{W}\!\!\!\!dx \int_{0}^{W}\!\!\!\!dy \,\,\nonumber\\
&&\times\,{\rm Im}\left( {{f_{nl}^{\sigma}}}^{\ast}
\frac{\partial}{\partial z} {f_{nl}^{\sigma}}
 + {{g_{nl}^{\sigma}}}^{\ast} \frac{\partial}{\partial z} {g_{nl}^{\sigma}}\right).
\label{QPcurrent}
\end{eqnarray}


\begin{figure}
{\includegraphics[width=0.8\columnwidth]{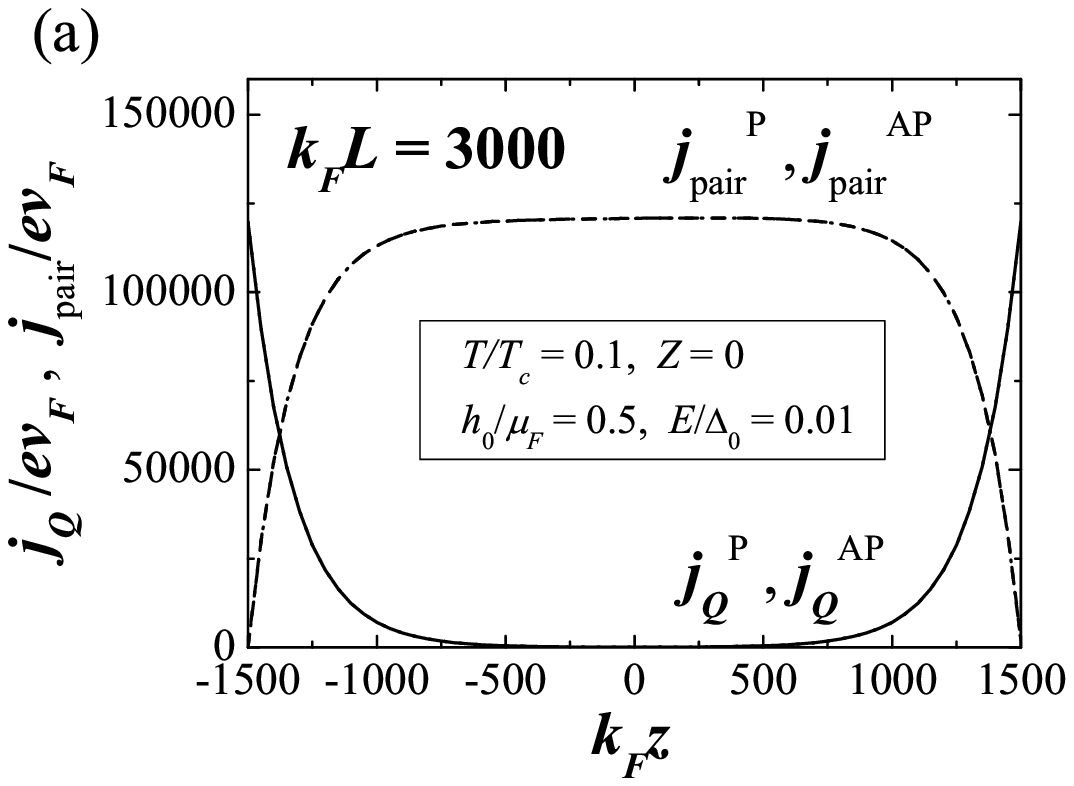}
\includegraphics[width=0.8\columnwidth]{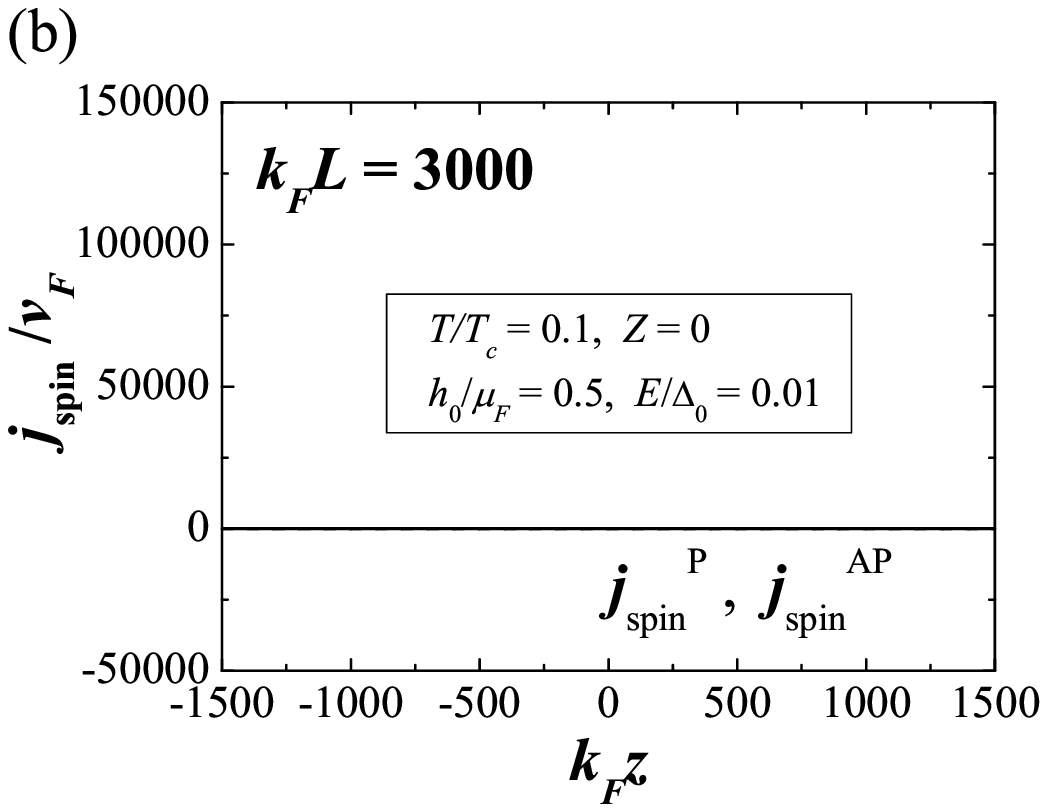}}
\caption{Spatial variation of the $z$ component of
  (a) the charge current density and (b) the spin current density
  in the SC with the thickness $L=3000/k_F$ is shown.
  $j_{Q}^{{\rm P(AP)}}$, $j_{\rm pair}^{{\rm P(AP)}}$ and
  $j_{\rm spin}^{{\rm P(AP)}}$
  are the QP current density, the supercurrent density and
  the spin current density
  in the parallel (anti-parallel) alignment, respectively.  
  We assume $\xi_{Q}(E=T=0) = 200/k_{F}$.}
\label{fig:3000}
\end{figure}

The right hand side of Eq. (\ref{contin}) corresponds to the gradient of supercurrent carried by Cooper pair ${\bf J}_{\rm pair}^{\sigma}$ defined as
\begin{eqnarray}
-\nabla\cdot {\bf J}_{\rm pair}^{\sigma} \equiv \frac{4e\Delta}{\hbar}\sum_{nl}{\rm Im}({f_{nl}^{\sigma}}^{\ast} g_{nl}^{\sigma}) ,
\end{eqnarray}
from which the $z$ component of ${\bf J}_{\rm pair}^{\sigma}$ per area $j_{\rm pair}^{\sigma}$ is obtained as 
\begin{eqnarray}
j_{\rm pair}^{\sigma} (z) = -\frac{4e\Delta}{\hbar W^2}\sum_{nl}
\int_{0}^{W}\!\!\!\!dx \int_{0}^{W}\!\!\!\!dy\nonumber\\
\times\,\int_{-L/2}^{z}\!\!\!\!dz' \,\,{\rm Im} ({{f_{nl}^{\sigma}}^{\ast}} {g_{nl}^{\sigma}}).  
\label{Scurrent}
\end{eqnarray}

The $z$ coordinate dependences of the QP current density $j_{Q} = j_{Q}^{\uparrow} + j_{Q}^{\downarrow}$ and the supercurrent $j_{\rm pair} = j_{\rm pair}^{\uparrow} + j_{\rm pair}^{\downarrow}$ in the case that the thickness of the SC
$L=3000/k_F$ are shown in Fig. \ref{fig:3000}(a).  
\begin{figure}
\includegraphics[width=\columnwidth]{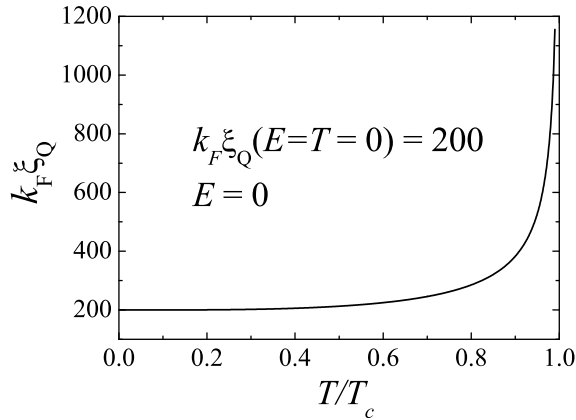}
\caption{The temperature dependence of the QP current
  penetration depth $\xi_Q$ for $E=0$.
    We assume $\xi_{Q}(E=T=0) = 200/k_{F}$.}
\label{fig:xiQ}
\end{figure}
We find that $j_{Q}$ decays from the interfaces of the FM1/SC and the SC/FM2 and becomes zero in the interior of the SC.  On the other hand, $j_{\rm pair}$ increases and becomes dominant in the interior of the SC to conserve the total current density.  In the energy region below the superconducting gap ($E<\Delta$) where the energy of the transverse mode $E_{nl}$ is smaller than Fermi energy $\mu_F$, the wave number $k^{\pm}_{nl}$ is expanded as
\begin{equation}
\begin{split}
k^{\pm}_{nl}
&\sim \frac{\sqrt{2m}}{\hbar} {\left( \mu_F \pm i\sqrt{\Delta^2 - E^2}\right) }^\frac{1}{2}\\
&\sim k_F \pm i \frac{1}{2\xi_Q}.  
\end{split}
\label{k-expansion}
\end{equation}
The imaginary part in Eq. (\ref{k-expansion}) gives the exponential decay term $\exp(-z/\xi_Q)$ in $j_{Q}$, where $\xi_Q$ is the penetration depth given by
\begin{equation}
\xi_Q = \frac{\hbar v_F}{2\sqrt{\Delta^2 - E^2}},
\end{equation}
where $v_F$ is the Fermi velocity.  
$\xi_Q$ has a strong temperature dependence shown in
Fig. \ref{fig:xiQ}.
  When the thickness of the SC is much larger than the penetration depth ($L \gg \xi_Q$) as in Fig. \ref{fig:3000}(a), $j_Q$ decays in the range of $\xi_Q$
from the interfaces.
Note that $\xi_Q$ is approximately equal to the clean-limit GL
coherence length $\xi(T)$ in the low energy regime:
$\xi_Q (E=0) \sim 1.2\thinspace\xi(T)$.\cite{BTK}

Next, we consider the spin transport in the SC.  The conservation law of the spin density $S=P_{\uparrow}-P_{\downarrow}$, where $P_{\sigma}=\sum_{nl}({|f_{nl}^{\sigma}|}^2+{|g_{nl}^{\sigma}|}^2)$, is derived from the BdG equation (\ref{eq:BdG}) and expressed as 
\begin{equation}
\frac{\partial S}{\partial t}+\nabla\cdot {\bf J}_{\rm spin} = 0 ,
\label{contin-spin}
\end{equation}
where ${\bf J}_{\rm spin} = {\bf J}_{P}^{\uparrow} - {\bf J}_{P}^{\downarrow}$ is the spin current density.  The $z$ component of ${\bf J}_{P}^{\sigma}$ per unit area $j_{P}^{\sigma}$ is written as
\begin{eqnarray}
j_{P}^{\sigma} (z) &=& \frac{\hbar}{mW^2}\sum_{nl}
\int_{0}^{W}\!\!\!\!dx \int_{0}^{W}\!\!\!\!dy \,\,\nonumber\\
&&\times\,{\rm Im}\left( {{f_{nl}^{\sigma}}}^{\ast}\frac{\partial}{\partial z} {f_{nl}^{\sigma}}
- {{g_{nl}^{\sigma}}}^{\ast} \frac{\partial}{\partial z} {g_{nl}^{\sigma}}\right).
\end{eqnarray}
Figure \ref{fig:3000}(b) shows the $z$ coordinate dependence of
the spin current density $j_{\rm spin} = j_{P}^{\uparrow} - j_{P}^{\downarrow}$ in the SC with the thickness $L=3000/k_F$.  
As shown in Fig. \ref{fig:3000}(b), no spin current flows through
the SC both in the parallel and in the anti-parallel alignments because the QP
current with spin changes to the supercurrent carried by Cooper
pairs with no spin.  This means that the spin injected
from FM1 does not reach to the FM2.  
As a result, the QP current density in the parallel alignment $j_{Q}^{\rm P}$
and that in the anti-parallel alignment $j_{Q}^{\rm AP}$ are almost the same as shown in Fig. \ref{fig:3000}(a).  

\begin{figure}
{\includegraphics[width=0.8\columnwidth]{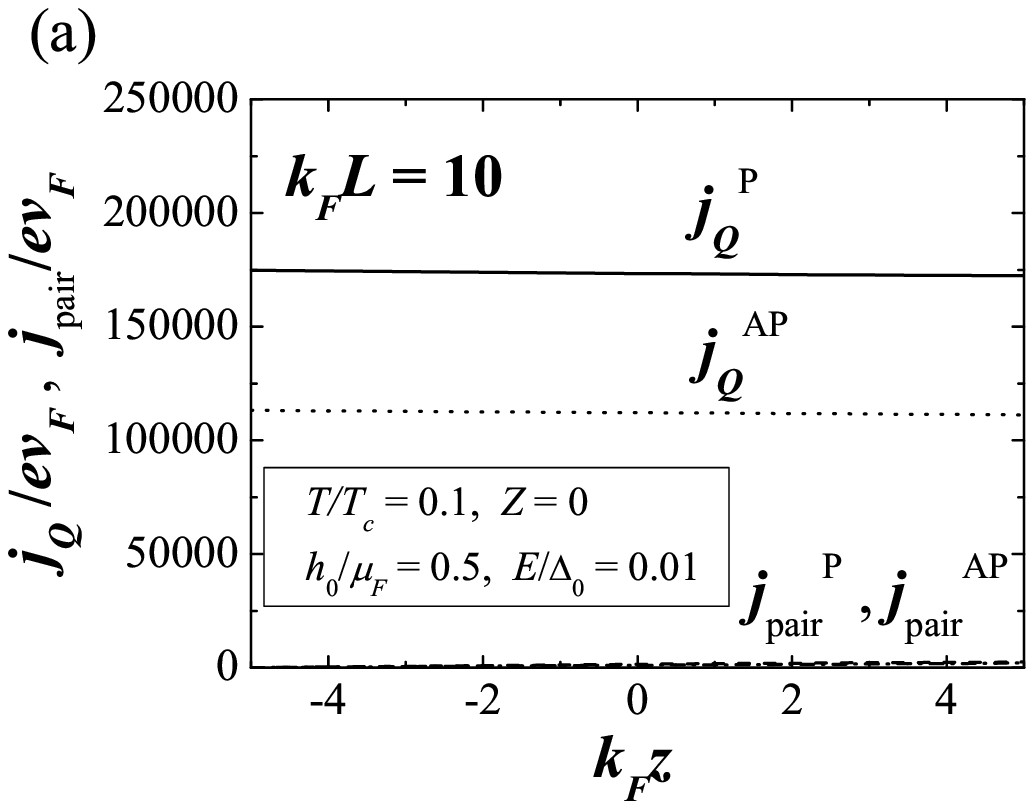}
\includegraphics[width=0.8\columnwidth]{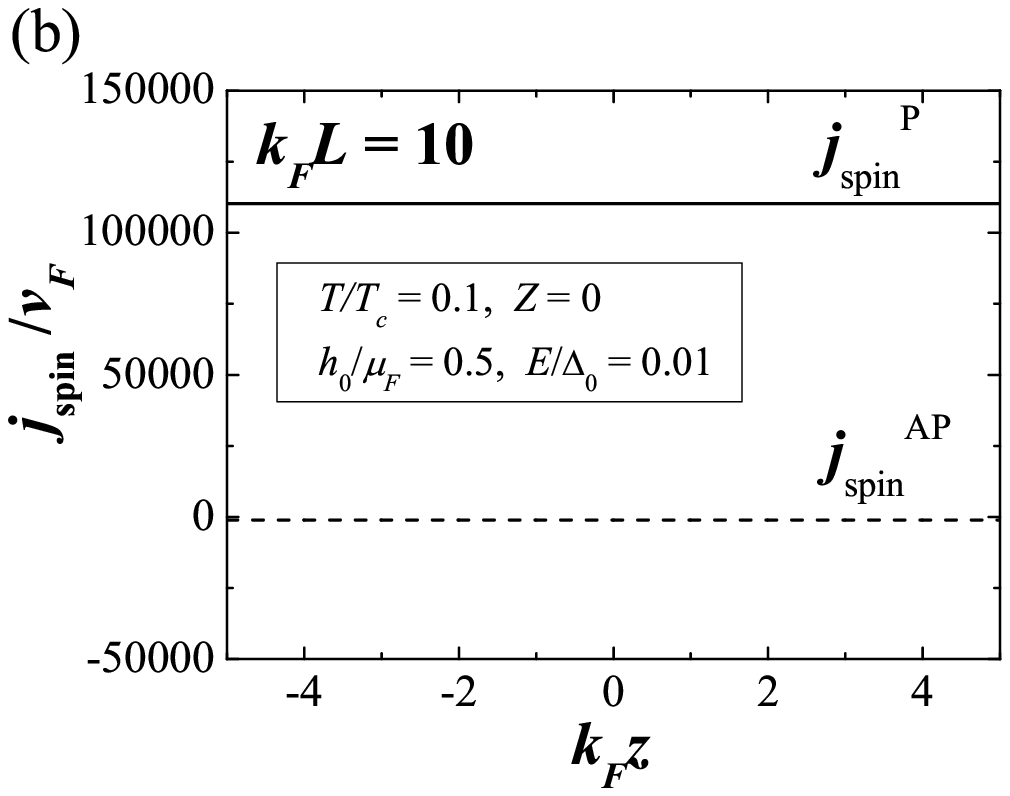}}
\caption{Spatial variation of the $z$ component of
  (a) the charge current density and (b) the spin current density
  in the SC with the thickness $L=10/k_F$ is shown.
  The parameters are the
  same as in Fig. \protect\ref{fig:3000}.}
\label{fig:10}
\end{figure}

The $z$ coordinate dependence of the charge current in the case that the
thickness of the SC is much smaller than the penetration depth ($L \ll \xi_Q$)
is shown in Fig. \ref{fig:10}(a), where the QP current density $j_{Q}$ is almost constant and the supercurrent density $j_{\rm pair}$ is nearly zero in the SC.  As shown in Fig. \ref{fig:10}(b), the value of the spin current
in the parallel alignment is larger than that in the anti-parallel alignment
because the value of QP current with up-spin
is much larger than that with down-spin in the parallel alignment,
whereas the value of QP current with up-spin is equal to
that with down-spin in the anti-parallel alignment.  
This means that the spin injected from the FM1 is transferred to the FM2
and therefore the value of the QP
current density strongly depends on the relative orientation of the FM's
magnetizations shown in Fig. \ref{fig:10}(a).

From the above discussion, the result shown in Fig. \ref{fig:Thick} is
understood as follows.  In the SC, the QP current with spin decreases
exponentially and changes to the supercurrent carried by Cooper pairs
with no spin in the range of $\xi_Q$ from the interfaces.  As a result,
it becomes difficult to transfer the spin from the FM1
to the FM2 and the MR decreases with increasing the thickness
of the SC.  The finite MR in the region of large $L$ is due to the
QP's with energy above the superconducting gap ($E>\Delta$).


\begin{figure}
{\includegraphics[width=0.8\columnwidth]{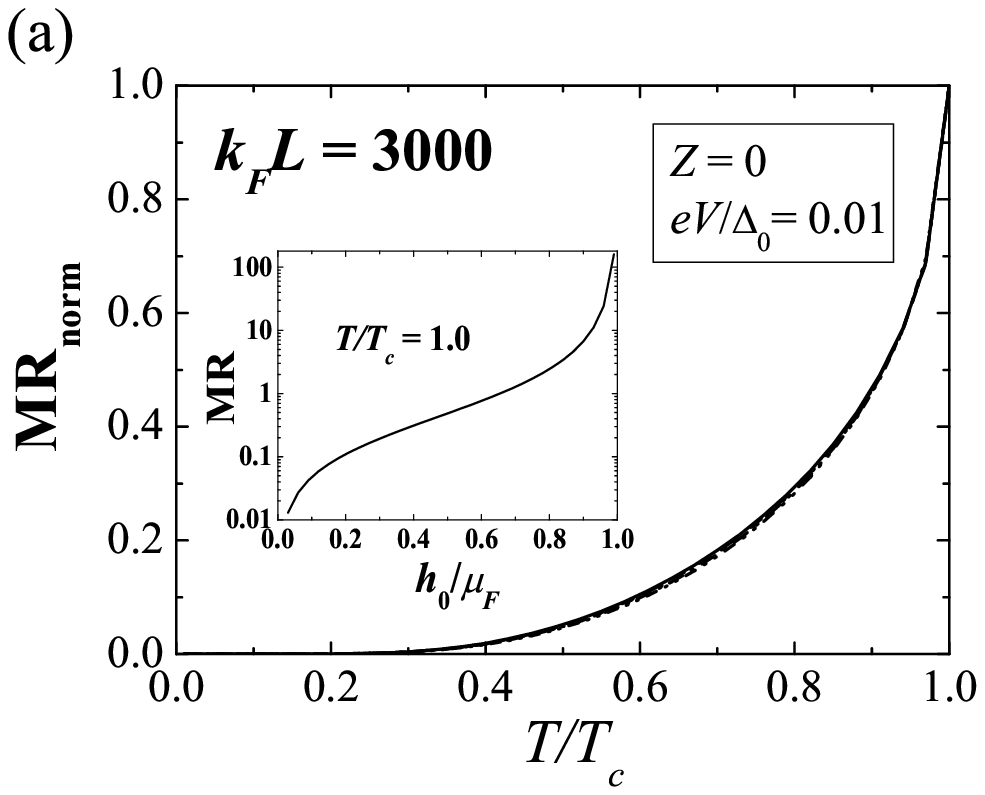}
\includegraphics[width=0.8\columnwidth]{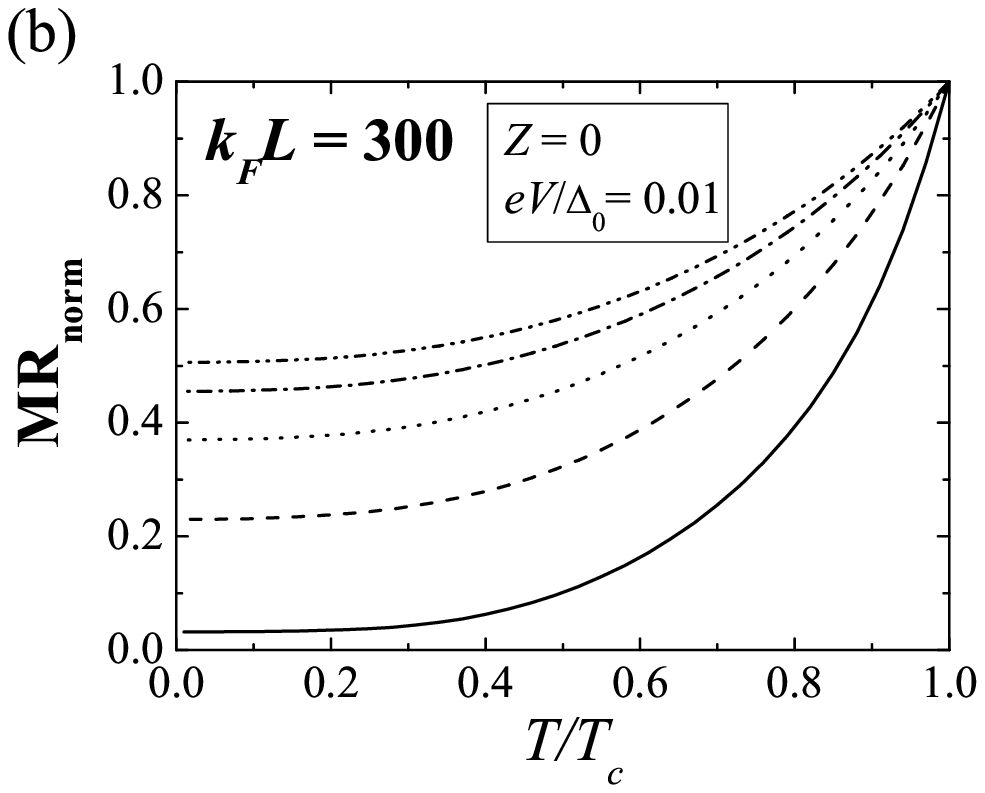}}
\caption{MR normalized by the value at $T_c$ (${\rm MR_{norm}}$)
  is plotted as a function of
  temperature. The thickness of the SC is taken to be $k_F L = 3000$
  and 300 in panels (a) and (b), respectively.
  From bottom to top, the exchange field $h_0 /\mu_F$ = 0.1, 0.3,
  0.5, 0.7, and 0.9.  The MR at $T=T_c$ is plotted as a function of the
  exchange field $h_0$ in the inset of panel (a).  
  We assume $\xi_{Q}(E=T=0) = 200/k_{F}$.}
\label{fig:MR_temp_pol}
\end{figure}
The temperature dependences of
the MR normalized by the value at $T_c$ (${\rm MR_{norm}}$)
for several values of the exchange field $h_0$
in the case of $L = 3000/k_F$ and $L = 300/k_F$ are shown in
Figs. \ref{fig:MR_temp_pol}(a) and \ref{fig:MR_temp_pol}(b), respectively.  
The ${\rm MR_{norm}}$ decreases with decreasing temperature because
the number of
electrons and holes with the energy $E>\Delta$ which contribute to the
${\rm MR_{norm}}$ decreases with decreasing temperature.
  At low temperatures, electrons and holes
mainly distribute in the energy region $E<\Delta$.  When the
thickness of the SC is much larger than the penetration depth
(Fig. \ref{fig:MR_temp_pol}(a)),
electrons and holes with energy $E<\Delta$ injected to the SC do not
contribute to
the ${\rm MR_{norm}}$ because the QP current changes to the supercurrent in the SC by the Andreev reflection and the QP transmission from the FM1 to
the FM2 does not occur.  As a result, the ${\rm MR_{norm}}$ becomes
zero at low temperatures $T/T_c \lesssim 0.4$.  
On the other hand, when the thickness of the SC is
comparable to the penetration depth
(Fig. \ref{fig:MR_temp_pol}(b)), the QP transmission by the Andreev reflection in the energy region $E<\Delta$ occurs
and therefore the finite ${\rm MR_{norm}}$ remains even at low temperatures.

\begin{figure}
{\includegraphics[width=0.8\columnwidth]{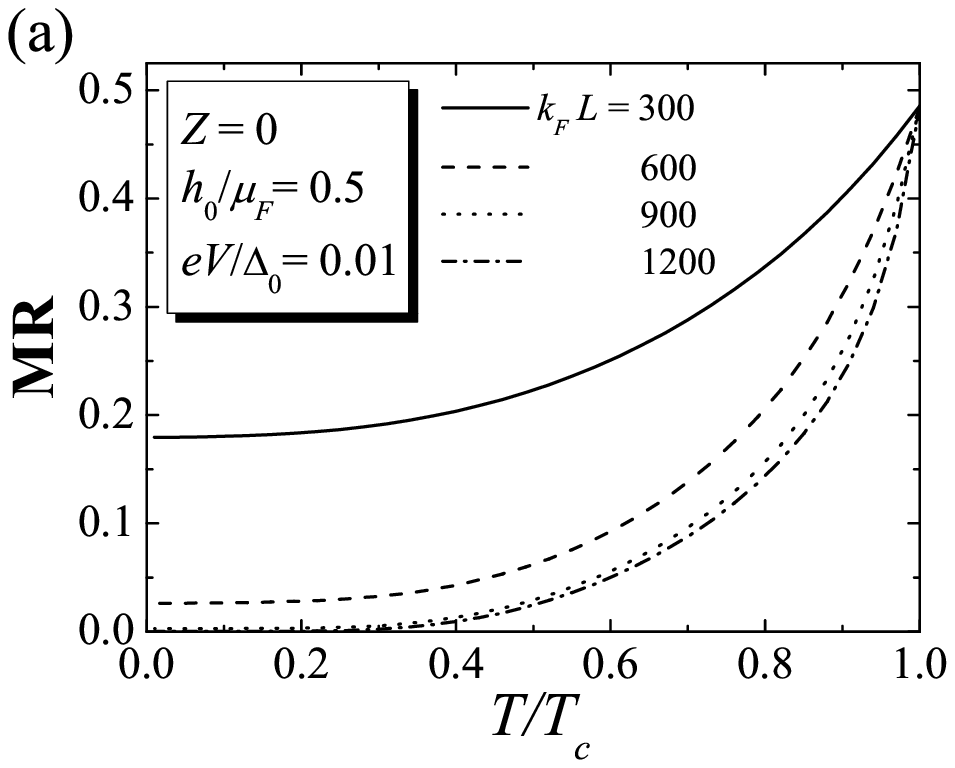}
\includegraphics[width=0.8\columnwidth]{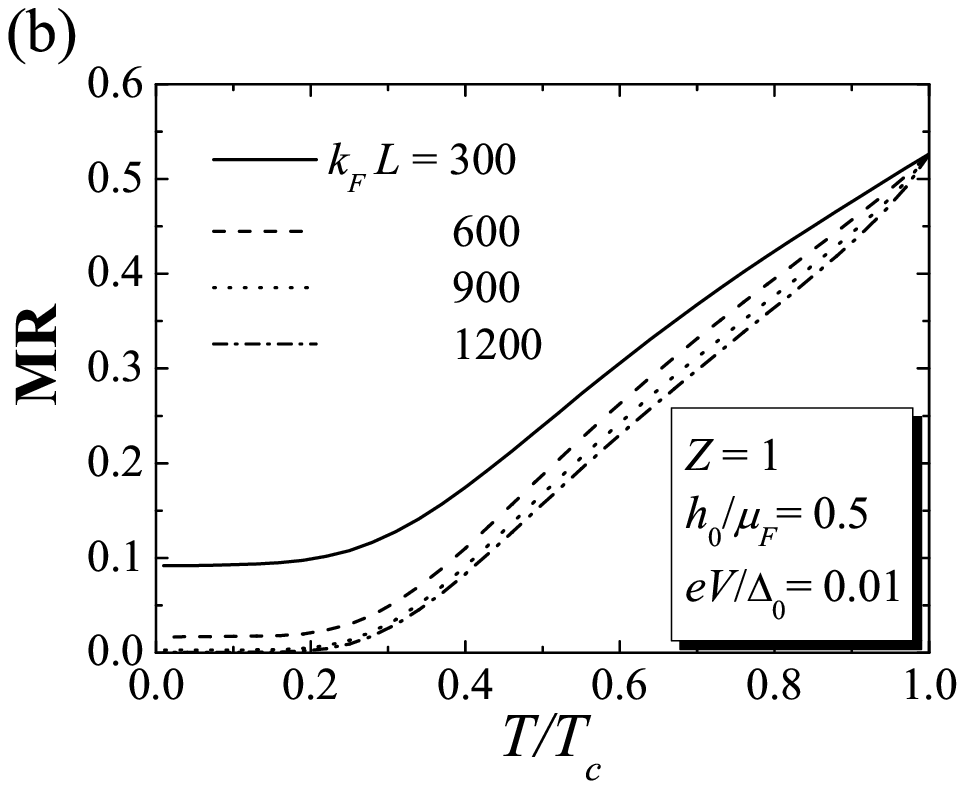}
\includegraphics[width=0.8\columnwidth]{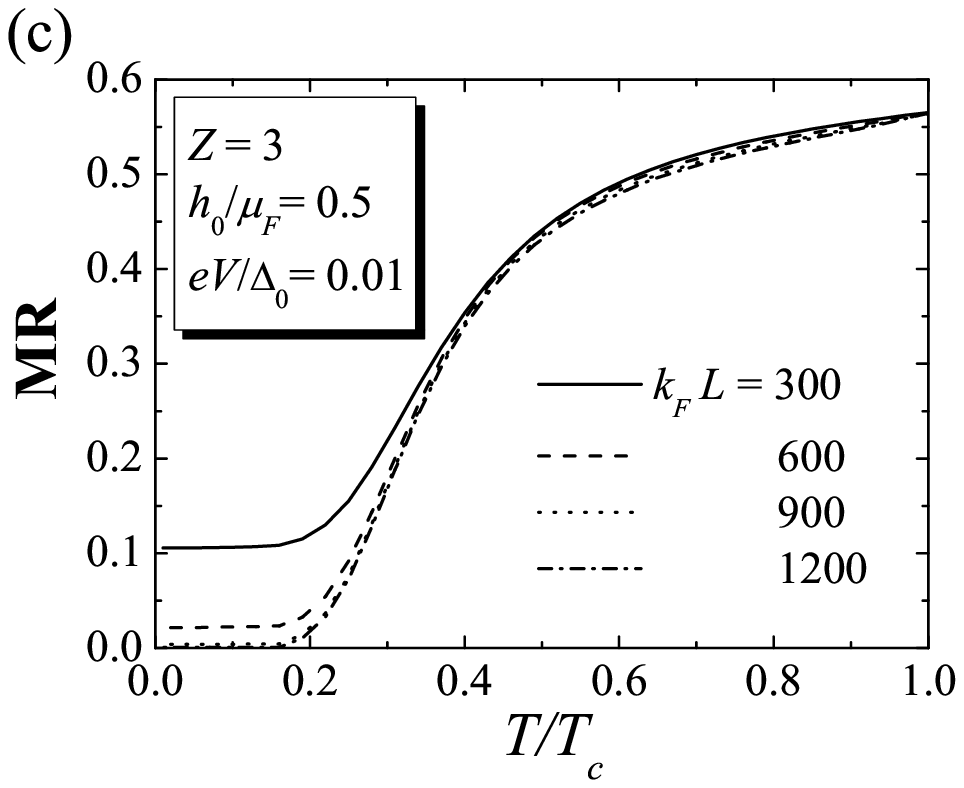}}
\caption{MR for the system with a several value of interfacial barrier is
  plotted as a function of temperature $T/T_c$.
  The strength of the interfacial  barrier is $Z=0$, $1$ and $3$
  for panels (a), (b) and (c) respectively.
  From top to bottom the thickness of the SC is
  $k_F L=$ 300, 600, 900, 1200.  
  We assume $\xi_{Q}(E=T=0) = 200/k_{F}$.}
\label{fig:MR_temp_finite_p05}
\end{figure}
Figures \ref{fig:MR_temp_finite_p05}(a)-(c) show the temperature
dependence of the MR in the cases of $Z=0$, $1$ and $3$, respectively,
for the several values of $L$.  
The temperature dependence of the MR in the case of the transparent interfacial
barrier (Fig. \ref{fig:MR_temp_finite_p05}(a)) is explained by the same way
as in Fig. \ref{fig:MR_temp_pol}.  
In the case of the finite interfacial barrier
(Figs. \ref{fig:MR_temp_finite_p05}(b) and \ref{fig:MR_temp_finite_p05}(c)),
for all values of $L$, the normal reflection mainly occurs
especially in the energy region below the superconducting gap ($E<\Delta$)
because of the scattering at the interfaces.  
Therefore, the main contribution to the MR at temperature
$0.3 \lesssim T/T_c \leq 1$ comes from the QP's transmission
in the energy region $E>\Delta$, whose probability is independent of $L$.  
As a result, the differences in the magnitude of the
MR for the different values of $L$ become smaller
especially for temperature $0.3 \lesssim T/T_c \leq 1$.

\section{Comparison with experiment}
\begin{figure}
\includegraphics[width=\columnwidth]{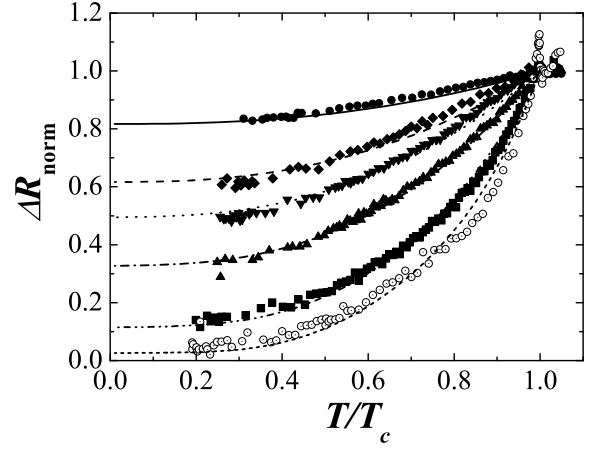}
\caption{MR is plotted as a function of temperature $T/T_c$.
  The solid curves show theoretical results for the thickness of the SC,
  $L=$30, 40, 50, 60, 80, and 100 nm from top to bottom,
  where $k_F$ is taken to be $1\,{\rm \r{A}^{-1}}$ for Nb.  
  The symbols show the experimental results by Gu {\it et al.}\cite{gu}
  for the thickness of the Nb,
  $t_{Nb}$=30, 40, 50, 60, 80, and 100 nm from top to bottom.}
\label{fig:pratt}
\end{figure}
Let us compare our theory with recent experimental results
in Py/Nb/Py structures measured by Gu {\it et al.}.\cite{gu}  
The mean free path in the Nb film $l \sim 6$ nm\cite{gu}
is much smaller than the clean-limit coherence length
$\xi_0 \sim 40$ nm\cite{Tinkham} and therefore the Nb film is in the
diffusive regime.  
In order to analyze the experimental results in the dirty Nb film,
we need to extend the theory in the ballistic case to that
in the diffusive case.  
The diffusive effect on the Andreev reflection is incorporated into
our theory by replacing the penetration depth $\xi_Q$ in the ballistic theory
with the penetration depth in the dirty-limit $\xi_Q^{D}$.  
Thus, the value of $\xi_Q(E=0)$ at $T=0$
obtained by fitting the experimental data is
interpreted as the dirty-limit penetration depth
$\xi_Q^{D}(E=0) \sim 1.2\sqrt{l/\xi_0}\,\xi_Q(E=0)$\cite{xiQD} at $T=0$.  
Figure \ref{fig:pratt} shows the excess resistance
$\Delta R = R_{\rm AP}-R_{\rm P}$
normalized by the value at (in the experiment, $T$ slightly above) $T_c$
($\Delta R_{\rm norm}$) as a function of temperature.  
The solid curves indicate the calculated results and the symbols indicate
the experimental ones.\cite{gu}  
By fitting the calculated values to those of the experimental data,
we obtain $\xi_{Q}^{D}(E=T=0)$ = 46, 36, 36, 33, 30, and 27 nm
for the curves of $L$=30, 40, 50, 60, 80, and 100 nm, respectively,
where $k_F$ is taken to be $1\,{\rm \r{A}^{-1}}$ for Nb.\cite{ashcroft}  
The values of the penetration depth $\xi_{Q}^{D}(E=T=0)$
estimated by our theory become larger for the smaller thickness of the Nb film
and are larger than the dirty-limit penetration depth in a bulk Nb
$\sim \sqrt{\xi_0 l}$ = 16.2 nm.  
This indicates that $\Delta$ in the Nb film is reduced compared to that
in a bulk Nb.  The suppression of $\Delta$ is due to the proximity effect.  
Actually, the height of the realistic superconducting gap depends on
the position $z$ in the Nb film by the proximity effect.  
Here we interpret the value of $\Delta$
as the averaged value of the realistic superconducting gap
with respect to $z$ in the Nb film.  
Gu {\it et al.} have obtained the dirty-limit penetration depth
$\xi_{Q}^D(E=T=0) \sim 16.5$ nm
by assuming that $\Delta R_{\rm norm}$ is proportional to
$\exp(-t_{Nb}/\xi_{Q}^D)$, where $t_{Nb}$ is the thickness of the Nb.  
This value of the penetration depth is comparable to
those estimated by our theory.

Although we neglect the effect of spin relaxation on the MR in our theory,
this assumption is justified as follows.  
In the Nb film in the Py/Nb/Py structure,
the spin diffusion length $\lambda_s$ is about $50$ nm,\cite{gu}
while the penetration depth $\xi_Q(T=0)$
is about $27 \sim 46$ nm at low energy.  
As seen in Fig. \ref{fig:xiQ}, $\xi_Q(T)$ is almost
constant at low temperatures and shows the divergent behavior only near $T_c$,
indicating that $\xi_Q$ is smaller than $\lambda_s$ in most of the temperature
range below $T_c$.  Therefore, the effect of the QP current penetration is
dominated for the MR and the spin relaxation effect on the MR is neglected
except in the close vicinity to $T=T_c$.

\section{Conclusion}\label{sec:conclustion}
The magnetoresistance in the ferromagnet/superconductor/ferromagnet double
junction system is studied theoretically.  The
dependences of the magnetoresistance on the thickness of the superconductor,
temperature, the exchange field of ferromagnets and the height of the
interfacial barriers are understood by considering
the Andreev reflection of spin-polarized current.  
Our theory shows good agreement with recent experimental results.
\acknowledgments
The authors thank W. P. Pratt, Jr. for informing them their experimental data
prior to publication.  
This work is supported by a Grant-in-Aid from MEXT and CREST of Japan.
H.~I. is supported by Encouragement of Young Scientists for MEXT.


\begin{references}
\bibitem{maekawaBOOK} {\it Spin Dependent Transport in Magnetic Nanostructures}
edited by S. Maekawa and T. Shinjo (Taylor and Francis, London and New York, 2002).

\bibitem{meservey} R. Meservey and P.M. Tedrow,
Phys. Rep. {\bf 238}, 173 (1994).  

\bibitem{vasko} V.A. Vas'ko, V.A. Larkin, P.A. Kraus, K.R. Nikolaev,
D.E. Grupp, C.A. Nordman, and A.M. Goldman,
Phys. Rev. Lett. {\bf 78}, 1134 (1997).  

\bibitem{dong} Z.W. Dong, R. Ramesh, T. Venkatesan, Mark Johnson, Z.Y. Chen,
S.P. Pai, V. Talyansky, R.P. Sharma, R. Shreekala, C.J. Lobb, and R.L. Greene,
Appl. Phys. Lett. {\bf 71}, 1718 (1997). 

\bibitem{yeh} N.-C. Yeh, R.P. Vasquez, C.C. Fu, A.V. Samoilov, Y. Li,
and K. Vakili,
Phys. Rev. B {\bf 60}, 10522 (1999).  

\bibitem{liu} J.Z. Liu, T. Nojima, T. Nishizaki and N. Kobayashi,
Physica C, {\bf 357-360}, 1614 (2001).  

\bibitem{chen} C.D. Chen, W. Kuo, D.S. Chung, J.H. Shyu, and C.S. Wu,
Phys. Rev. Lett. {\bf 88}, 047004 (2002).  

\bibitem{takahashiPRL} S. Takahashi, H. Imamura, and S. Maekawa,
Phys. Rev. Lett. {\bf 82}, 3911 (1999). 

\bibitem{yamashita} T. Yamashita, S. Takahashi, H. Imamura, S. Maekawa, 
Phys. Rev. B {\bf 65}, 172509 (2002).

\bibitem{takahashiJMMM} S. Takahashi, T. Yamashita, H. Imamura, S. Maekawa, 
J. Magn. Magn. Mater. {\bf 240}, 100 (2002).

\bibitem{takahashiHOLL} S. Takahashi and S. Maekawa,
Phys. Rev. Lett. {\bf 88}, 116601 (2002).

\bibitem{de_jong} M.J.M. de Jong and C.W.J. Beenakker,
Phys. Rev. Lett. {\bf 74}, 1657 (1995).  

\bibitem{kikuchi} K. Kikuchi, H. Imamura, S. Takahashi, and S. Maekawa,
Phys. Rev. B {\bf 65}, 020508(R) (2001).  

\bibitem{hima} H. Imamura, K. Kikuchi, S. Takahashi, and S. Maekawa,
J. Appl. Phys. {\bf 91}, 172509 (2002).  

\bibitem{zhu} J.-X. Zhu, B. Friedman, and C.S. Ting,
Phys. Rev. B {\bf 59}, 9558 (1998).  

\bibitem{igor} I. \v{Z}uti\'{c} and O.T. Valls,
Phys. Rev. B {\bf 60}, 6320 (1999).  

\bibitem{kashiwaya} S. Kashiwaya, Y. Tanaka, N. Yoshida, and M.R. Beasley,
Phys. Rev. B {\bf 60}, 3572 (1999).  

\bibitem{soulen} R.J. Soulen Jr., J.M. Byers, M.S. Osofsky, B. Nadgorny,
T. Ambrose, S.F. Cheng, P.R. Broussard, C.T. Tanaka, J. Nowak, J.S. Moodera,
A. Barry, and J.M.D. Coey,
Science {\bf 282}, 85 (1998).  

\bibitem{soulen2} R.J. Soulen, Jr., M.S. Osofsky, B. Nadgorny, T. Ambrose,
P. Broussard, S.F. Cheng, J. Byers, C.T. Tanaka, J. Nowack, J.S. Moodera,
G. Laprade, A. Barry, and M.D. Coey,
J. Appl. Phys. {\bf 85}, 4589 (1999).

\bibitem{ji} Y. Ji, G.J. Strijkers, F.Y. Yang, C.L. Chien, J.M. Byers,
A. Anguelouch, Gang Xiao, and A. Gupta,
Phys. Rev. Lett. {\bf 86}, 5585 (2001).

\bibitem{strijkers} G.J. Strijkers, Y. Ji, F.Y. Yang, C.L. Chien,
and J.M. Byers,
Phys. Rev. B {\bf 63}, 104510 (2001).

\bibitem{shashi} S.K. Upadhyay, A. Palanisami, R.N. Louie, and R.A. Buhrman,
Phys. Rev. Lett. {\bf 81}, 3247 (1998).

\bibitem{andreev} A.F. Andreev, Sov. Phys. JETP {\bf 19}, 1228 (1964).  

\bibitem{GL} V.L. Ginzburg and L.D. Landau,
Zh. Eksperim. i Teor. Fiz. {\bf 20}, 1064 (1950).  
\bibitem{BTK} G.E. Blonder, M. Tinkham, and T.M. Klapwijk,
Phys. Rev. B {\bf 25}, 4515 (1982).  

\bibitem{gu} W.P. Pratt, Jr. ; private communication,
J.Y. Gu, J.A. Caballero, R.D. Slater, R. Loloee, and W.P. Pratt, Jr.,
to be published in Phys. Rev. B.  

\bibitem{BdG} P.G. de Gennes,
{\it Superconductivity of Metals and Alloys} (W. A. Benjamin, New York, 1966),
chap. 5.  

\bibitem{belzig} W. Belzig, Arne Brataas, Yu.V. Nazarov and Gerrit E.W. Bauer,
Phys. Rev. B {\bf 62}, 9726 (2000).

\bibitem{lambert} C.J. Lambert, J. Phys.: Condens. Matter {\bf 3}, 6579 (1991).

\bibitem{Tinkham} M. Tinkham,
{\it Introduction to Superconductivity} (McGraw Hill, New York, 1996).  

\bibitem{xiQD} 
We assume that the dirty-limit penetration depth is expressed as
$\xi_{Q}^{D}(E=0)\sim 1.2\,\xi^{D}(T)$ by analogy with the clean-limit case,
where $\xi^{D}(T)$ is the dirty-limit GL coherence length.  
Using the relation
$\xi^{D}(T) \sim 1.2\sqrt{l/\xi_0}\,\xi (T)$,\cite{Tinkham}
the dirty-limit penetration depth can be expressed as
$\xi_Q^{D}(E=0) \sim 1.2\sqrt{l/\xi_0}\,\xi_Q(E=0)$.  

\bibitem{ashcroft} N. Ashcroft and N. Mermin,
{\it Solid State Physics} (Saunders College Publishing, New York, 1976).
\end{references}
\end{document}